# Neko: A Modern, Portable, and Scalable Framework for High-Fidelity Computational Fluid Dynamics

Niclas Jansson
KTH Royal Institute of Technology
Stockholm, Sweden
njansson@kth.se

Martin Karp
KTH Royal Institute of Technology
Stockholm, Sweden
makarp@kth.se

Artur Podobas
KTH Royal Institute of Technology
Stockholm, Sweden
podobas@kth.se

Stefano Markidis
KTH Royal Institute of Technology
Stockholm, Sweden
markidis@kth.se

Philipp Schlatter
KTH Royal Institute of Technology
Stockholm, Sweden
pschlatt@mech.kth.se

## ABSTRACT

Recent trends and advancement in including more diverse and heterogeneous hardware in High-Performance Computing is challenging software developers in their pursuit for good performance and numerical stability. The well-known maxim "software outlives hardware" may no longer necessarily hold true, and developers are today forced to re-factor their codebases to leverage these powerful new systems. CFD is one of the many application domains affected. In this paper, we present Neko, a portable framework for high-order spectral element flow simulations. Unlike prior works, Neko adopts a modern object-oriented approach, allowing multi-tier abstractions of the solver stack and facilitating hardware backends ranging from general-purpose processors down to exotic vector processors and FPGAs. We show that Neko's performance and accuracy are comparable to NekRS, and thus on-par with Nek5000's successor on modern CPU machines. Furthermore, we develop a performance model, which we use to discuss challenges and opportunities for high-order solvers on emerging hardware.

## 1 INTRODUCTION

Computational fluid dynamics (CFD) is at the heart of modern engineering and an indispensable tool for areas such as automotive, aerospace, energy, weather and climate. A particularly important area here is turbulence, as about 10% of the energy use in the world is spent overcoming turbulent friction [23]. Improvements in this area have thus a clear environmental and societal impact. With a virtually unbounded need of computational resources, CFD is a natural driver for exascale computing [56], and CFD HPC applications and codes are not uncommon Gordon Bell Prize finalists [2] and winners [48, 53].

One of the most common ways of simulating large scale incompressible flows on HPC systems has been using the Nek5000 application. Nek5000 introduced in the mid-nineties by Fischer et al. [48] tracing its roots to the (even older) NEKTON 2.0 [15, 24, 52] framework and has since been a steady and reliable companion in the CFD community because of its reliability and scalability. It is one of the few applications that can scale to a million cores. Nek5000 obtained the Gordon Prize Award in 1999 [60]. Without a doubt, for many, Nek5000 has symbolized (and still does) high-quality numerics brought together under the umbrella of a standard (Fortran) language, where only necessary components are integrated to avoid code-bloat or unnecessary dependencies– at least until recently.

Today, HPC is approaching an exascale-level of performance [12]. Coincidentally, Moore's law [55] – upon which HPC has relied much of its improvement – is diminishing and is expected soon to end [33]. These observations, in turn, have forced system architectures and manufacturers to more aggressively incorporate elements of heterogeneity into the systems. Heterogeneity in terms of computing allows for specialization– the opposite of the general-purpose off-the-shelf trend that we have grown used to. Nowhere is this trend as clear as the TOP500 list [57], where the five top entries are either specially built for the purpose (A64FX [65] and Sunway Taihulight [11]) or contain Graphics Processing Units (GPUs). This shift in trend puts new constraints and requirements on legacy software, such as Nek5000, which now need to undergo significant refactoring and change to keep up with changes in architectures. Once praised, Nek5000 features – such as the lack of dynamic memory and the extensive use of Fortran77 common blocks – now harm more than help.

Today, three efforts have answered Nek5000's call for aid. The first effort relies on augmenting the original Nek5000 with OpenACC [8] in order to cater to the need for accelerators. Unfortunately, Nek5000 was designed in a time where accelerators did not yet exist, and adopting OpenACC into such a codebase is extremely inflexible and does not address the issue with the underlying legacy features. Furthermore, OpenACC caters only to a single type of accelerator, namely Nvidia GPUs, and ignores emerging technology such as Field-Programmable Gate Arrays (FPGAs) or Vector engines. The second effort is NekRS [16, 41], which is a complete rewrite of the framework. Here, users depend on Just In Time compilation (JIT) and the code is written in a Domain-Specific Language (DSL) that relies on OCCA [40] to automatically generate kernels suitable for different computational platforms. Unfortunately, NekRS in its current form relies heavily on the success (or failure) of OCCA and focuses exclusively on an offload model (which we argue might not be the best approach). This third effort – which is also the effort we introduce in this paper – is Neko.

Neko aspires to be as performant, easy-to-use, and numerically stable as Nek5000, but with the added benefits of dynamic memory, a (multi-)modular codebase in modern Fortran 2008, and support for many modern accelerator backends. Neko was built with accelerators in mind and supports offloading subsets of the computation to GPUs and FPGAs, or native execution on Vector processors. Finally – unlike prior work – Neko is built bottom-up on a performance model, which guides and steers performance aspects of the development process.

In short, we claim the following contributions:

(1) We introduce Neko, a modernized framework for high fidelity CFD simulations on modern supercomputers,
(2) We reveal inner implementation details and design decision in how we achieve modularity, numerical stability, and high-performance,
(3) We empirically compare and contrast Neko against both NekRS and Nek5000 in terms of absolute performance (strong scaling) and numerical accuracy (verification) on well-established benchmarks,
(4) We develop a performance model, which we validate and use to project performance on-to future (up-coming) architectures in order to reveal future challenges and opportunities for large scale high-performance CFD simulations.

## 2 FINDING A SUITABLE ABSTRACTION

When designing both a flexible and maintainable framework for computational science, a major issue is to find the right level of abstraction. Allowing for abstractions throughout the entire framework might degrade performance in low-level kernels while keeping the abstractions only at the top levels results in a codebase with many specialized kernels and a high maintenance cost.

The underlying numerical method also mandates at which level and what kind of abstraction would be possible. The weak form used in the Finite Element Method (FEM) is an excellent example of how to keep the abstractions at the top level, for example, given a bilinear and linear forms $a$, $L$ and a function space $V$, we formulate the abstract problem as, find $u \in V$ such that,

$$a(u,v) = L(v) \quad \forall v \in V. \quad (1)$$

Given a particular discretisation of a domain, we make an ansatz for the solution $u$ as:

$$u = \sum_{j=1}^{n} \xi_j \varphi_j, \quad (2)$$

where $\xi_j = u(n_j)$, $n$ is the number of nodes $n_j$ in the discretisation and $\varphi_j$ the finite element basis functions. Substituting (2) into (1) gives the discrete abstract problem,

$$\sum_{j=1}^{n} \xi_j a(\varphi_j, \varphi_i) = L(\varphi_i), \quad i = 1, \ldots, n, \quad (3)$$

with a corresponding discrete system $A\xi = b$, where $A_{ij} = a(\varphi_j, \varphi_i)$ and $b_i = L(\varphi_i)$. Thus, a FEM framework that provides a solver for $A\xi = b$ and an abstraction for computing $A_{ij}$ would be able to solve any problem that could be expressed as in (1). For higher-order methods, e.g. spectral elements, a similar abstraction can be achieved. Assume we want to solve Poisson's equation with homogeneous Dirichlet boundary conditions,

$$-\nabla^2 u = f \quad \text{in } \Omega, \quad (4)$$
$$u = 0 \quad \text{on } \partial\Omega. \quad (5)$$

The spectral element solution is based on the weak formulation, namely find $u \in V \subset H_0^1$ such that,

$$\int_\Omega \nabla u \nabla v d\Omega = \int_\Omega f v d\Omega \quad \forall v \in V. \quad (6)$$

We discretize $\Omega$ into a set of $E$ non-overlapping hexahedral elements $\Omega = \cup_{e=1}^{E} \Omega^e$ and define a piecewise polynomial approximation space $V^N$, with a tensor-product polynomial basis using one dimensional $N$-th order Legendre-Lagrange polynomials $l_i(\xi)$,

$$l_i(\xi) = \frac{N(N+1)^{-1}(1-\xi^2)L'_N(\xi_i)}{(\xi - \xi_i)L_N(\xi_i)} \quad \text{for } \xi \in [-1, 1] \quad (7)$$

with Gauss-Lobatto-Legendre (GLL) quadrature points $\xi_i$, and $N$-th order Legendre polynomials $L_N$. The discrete solution $u$ can then be expressed as a tensor product of the polynomials on the reference element,

$$u^e(\xi, \eta, \gamma) = \sum_{i,j,k}^{N} u_{ijk}^e l_i(\xi) l_j(\eta) l_k(\gamma), \quad (8)$$

where $\xi, \eta, \gamma \in [-1, 1]$ are the coordinates of the reference element. Applying this formulation, the discrete, bilinear form $a(u, v)$ of (6) can be expressed as,

$$a(u,v) = \sum_{e=1}^{E} (v^e)^T D^T G^e D u^e = \sum_{e=1}^{E} (v^e)^T A^e u^e, \quad (9)$$

where $G^e$ is the tensor comprising of the geometric factors for mapping to and from the reference element and $D$ the local derivatives of the operand at the GLL points.

We have now arrived at a similar abstraction level as for the discrete abstract problem showed in (3). However, assembling either the local element matrix $A^E$ or the full stiffness matrix $A$ is prohibitively expensive in high-order methods. Therefore, a key to achieve good performance in the Spectral Element Method (SEM) is to consider a matrix-free formulation, where one always works with the unassembled matrix $A_L = \text{diag}\{A^1, A^2, \ldots, A^E\}$. Each degree of freedom in the discrete solution $u$ is assigned a unique global number. To ensure continuity of functions on the element level $u_{ijk}^e$, we define a Boolean gather matrix $Q^T$, mapping from local to global (unique) numbers in $u$. A corresponding scatter matrix is given by $Q$ such that $u_L = \{u^e\}_{e=1}^E$. With the gather-scatter operations the discrete bilinear form (9) becomes

$$a(u,v) = \sum_{e=1}^{E} (v^e)^T A^e u^e = (Qv)^T A_L Qu = v^T Au. \quad (10)$$

As described in [10], $Q$ and $Q^T$ are never formed explicitly, only the action $QQ^T$ is used as a single gather-scatter operation. For example, the matrix-vector product $w = Au$, becomes $w_L = QQ^T A_L u_L$ in the matrix-free formulation. Thus, with an efficient way of formulating a matrix-vector product $Ax$, the final component necessary for a generic spectral element framework using matrix-free Krylov-subspace methods is established.

## 2.1 Programmability

With the appropriate mathematical abstraction offered by SEM, the question is how to leverage it in an application. One approach is to adopt a Domain-Specific Language (DSL) that can express a given equation and generate low-level code for computing, e.g. $Ax$. FEniCS [37] is an excellent example of this approach, with a DSL used to express the bilinear and linear forms in (1) from which code is generated to compute $A$ and $b$ in $A\xi = b$.

Although very flexible and general, a variational-form DSL, like the one in FEniCS, comes with a high maintenance price, both for ensuring correctness in translating from DSL to code and ensuring support and performance on various platforms. Furthermore, one also needs to consider how often the code generation step is executed. For a general-purpose computational framework, like FEniCS, very often, but for e.g. a flow solver, the DSL would only be used during prototyping and seldom, if at all, in production.

The DSL or abstraction can be moved from the variational form down to the actual kernel implementation for more problem-specific solvers. Prime examples are, Kokkos [13], a library consisting of a performance portable abstraction layer in C++ targeting both CPU and GPU architectures, the Oxford Parallel DSLs: OP2 [43] and OPS [51] libraries and abstraction layer for automatic parallelisation of unstructured and multi-block structured algorithms and OCCA [40], a library and DSL, or rather a kernel language, with just-in-time compilation for various backends. Both libraries have successfully been used in various CFD codes, e.g. SPARC [28], which uses Kokkos and Nek5000's successor NekRS built on top of OCCA, and OPS is the backbone of the scalable OpenSBLI [9] framework. However, there may be drawbacks to these approaches as well.

Employing either of the abstraction layers in an application requires non-trivial integration at a deep level, via API calls, e.g. OP2, specific data types, e.g. Kokkos-arrays or via DSL snippets for the OCCA and OPS libraries. Abstraction requiring specific data types will implicitly enforce consumer applications to use the same programming language or end up with a mixed-language codebase that might become a maintenance nightmare. Similarly, API based abstraction allows for better cross-language interoperability, but unless data types can seamlessly move between the simulation code and the abstraction, one will end up in the same mixed-language issues as before. With a DSL approach, the impact is less severe and can be limited only to affect specific computational kernels. However, for all these approaches, there is always a question about portability and sustainability.

The success and longevity of Nek5000 (more than 30 years) can in part be attributed to the design choice of using a standardised programming language, Fortran 77, in this case. Thus, as long as any future platform has a standard-conforming Fortran 77 compiler, we would also have a working Nek5000 solver on that platform. Hence, portability and long term sustainability could be the unfortunate Achilles' heel of abstraction layers such as Kokkos and OCCA. Furthermore, it is also a question about programming models and their support on current and future platforms. Today, most heterogeneous systems with accelerators are programmed using a model that offloads computation from the host (CPU) to the accelerator (often a GPU). To benefit from these systems, applications must consider how the offloaded kernels are computed, and as necessary, how to optimise memory transfers between host and accelerator. What will happen if we suddenly get a new accelerator, which does not operate following the current offloading convention? An excellent example of this is the recent vector computer from NEC, SX-Aurora, a PCI express card called a Vector Engine (VE), that, unlike GPUs, preferably is used in native mode by executing the entire application on the accelerator (even communication parts).

Given the challenges of finding a portable programming abstraction, we argue that another option is to design application-specific abstraction layers to facilitate ease of use of different hardware-specific kernels and solvers for a particular programming model. Furthermore, an application-specific abstraction could also consider using small reproducers of kernels and finding suitable transformations in frameworks such as DaCe [3] before adding the final, tuned kernel back to the codebase. Is this not the same then, as using, e.g. OCCA and Kokkos from the start? No, this method removes dependencies introduced via these abstraction layers and potential portability issues on future platforms.

## 3 NEKO

Neko is a portable framework for high-order spectral element based simulations on hexahedral meshes, mainly focusing on incompressible flow simulations based on the proven numerical methods of Nek5000, and fast operator evaluations pioneered by Orszag [45]. The framework is implemented in Fortran 2008 and adopts a modern object-oriented approach, allowing multi-tier abstractions of the solver stack and facilitating various hardware backends.

Compared to Nek5000 and its successor NekRS, Neko is not simply a rewrite using modern Fortran and object orientation; the framework departs from the monolithic solver design (inherited from the static memory model of Fortran 77), allowing for multiple usage models. The simplest usage model is only to use the solver installed when building Neko. This binary is a generic solver requiring a mesh and an input deck to set up and run a case in contrast to Nek5000 (and to some extent NekRS), where the entire code base needs to be recompiled for each case to determine the length of static arrays and accommodate for user-defined functions, e.g. inflow conditions or post-processing routines.

On the contrary, Neko's solver is built on top of a library `libneko`. If a user can not set up a problem using built-in functionalities, e.g. inflow conditions, the library provides a set of callbacks that a user can define via a simple build script similar to Nek5000's `makenek`. A user can also directly work against Neko's library, using various components to develop a stand-alone SEM-based solver.

Table 1 gives a detailed list of Nek5000, NekRS and Neko's differences, and in the remainder of this section, we highlight certain key features of Neko, setting it apart from Nek5000's design philosophy.

## 3.1 Abstraction Layer

The multi-tier abstractions in Neko is realised using abstract Fortran types, with deferred implementations of required procedures. To allow for different formulations of a simulation's governing equations, Neko provides an abstract type `ax_t`, defining a matrix-vector product. The abstract type is shown in Listing 1 and requires any

| Feature | Neko | Nek5000 | NekRS |
|---|---|---|---|
| Dynamic Memory | Yes | No | Yes |
| Dynamic Polynomial degree | Yes | No | JIT |
| Unified Mesh Format | Yes | No | No |
| Multi-backend support | Yes | No | Yes |
| Modular design | Yes | No | Yes |
| Only standard library dependencies | Yes | Yes | No |
| Strict type-checking | Yes | No | Yes |
| Variable precision support | Yes | No | No |
| Dynamic choice of solvers | Yes | No | JIT |
| Multi compiler support | Yes | Yes | No |
| Native gather-scatter operator | Yes | No | No |

Table 1: Current features of the different solvers.

Listing 1: The abstract matrix-vector product type.

```
! Base type for a matrix-vector product providing Ax
type, abstract :: ax_t
 contains
   procedure(ax_compute), nopass, deferred :: compute
end type ax_t

! Abstract interface for computing Ax
abstract interface
   subroutine ax_compute(w, u, coef, msh, Xh)
     implicit none
     type(space_t), intent(inout) :: Xh
     type(mesh_t), intent(inout) :: msh
     type(coef_t), intent(inout) :: coef
     real(kind=dp), intent(inout) :: w(:,:,:,:)
     real(kind=dp), intent(inout) :: u(:,:,:,:)
   end subroutine ax_compute
end interface
```

derived, concrete type to provide an implementation of the deferred procedure compute that would return the action of multiplying the stiffness matrix of a given equation with a vector. For Poisson's equation, this corresponds to a routine for computing (9).

In typical object-oriented fashion, whenever a routine needs a matrix-vector product, it is always expressed as a call to compute on the abstract base type and never on the actual concrete implementation. Abstract types are all defined at the top level in the solver stack during initialisation and represent large, compute-intensive kernels, thus reducing overhead costs associated with the abstraction layer.

Furthermore, this abstraction also accommodates the possibility of providing tuned ax_t for specific hardware, only providing a particular implementation of compute in 1 without having to modify the entire solver stack. The ease of supporting various hardware backend is the key feature behind the performance portability of Neko on general-purpose processors down to exotic vector processors [29], and field-programmable gate arrays [32].

However, a portable matrix-vector multiplication backend is undoubtedly not enough to ensure performance portability. Therefore, an abstract type is used to describe a flow solver's common parts, with a deferred procedure for computing a time-step. Figure 1 illustrate this for a canonical simulation in Neko. Each case (case_t) is defined based on a mesh (mesh_t) together with an abstract solver (solver_t), later defined as an actual extended derived type based on the simulation parameters. Each of these abstract solvers contains a set of defined derived types necessary for a spectral element simulation, such as a function space (space_t), coefficients (coef_t) and various fields (field_t). Additionally, it contains further abstract types for defining matrix-vector products (ax_t) and gather-scatter $QQ^T$ kernels (gs_t). Each of these abstract types is associated with an actual implementation in an extended derived type (green boxes), allowing for hardware or programming model specific implementations, all interchangable at runtime. This way, Neko can accommodate both native and offloading type execution models without too much code duplication in the solver stack.

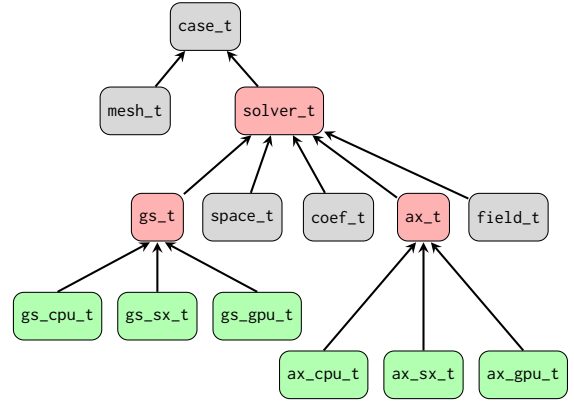

Figure 1: An illustration of a canonical flow case in Neko, with typical derived types (gray), abstract types (red) with actual implementation in extended derived types (green).

### 3.2 Gather-Scatter

As described in Section 2, all operations are performed in a matrix-free fashion for an efficient SEM implementation. Thus, the gather-scatter operator $QQ^T$ must be applied to all matrix-vector operations to ensure continuity across elements. Furthermore, since matrix-vector multiplication is at the core of an iterative Krylov solver, an efficient (and scalable) gather-scatter operation is crucial.

Both Nek5000 and NekRS are using the C library gslib [17, 38] for gather-scatter operations. The library is written as a generic sparse communication kernel, capable of performing various gather-scatter operations, e.g. addition or multiplication as a black-box, only requiring a list of shared (global) ids and a local to global mapping. To efficiently carry out the gather-scatter operation, gslib groups local indices in $u^e_{ijk}$ sharing the global id $\hat{\imath}$ in $u_{\hat{\imath}}$, stored in a consecutive list terminated with clear markers between each global id. How these chunks, related to an id $\hat{\imath}$, is processed depends on the underlying hardware, with various flavours of gslib for specific architectures see for example [47] for some of the GPU optimisations.

Compared to gslib, Neko provides native gather-scatter operations in the solver framework, implemented in Fortran, following the same abstract object-oriented design to provide portable gather-scatter kernels and therefore also ensuring portability of matrix-vector functions. Furthermore, with the implementation directly in the framework, the gather-scatter kernel can utilise information about the underlying computational mesh. Based on the topology, gather-scatter operations are scheduled to allow for overlapping communication with computation. Elements are classified as *local*, if all shared global ids $u_{\hat{\imath}}$ are owned by the same Processing Element (PE). If one or more ids are shared with another PE, the element is classified as *shared*. Based on the element classification, each global id is stored in a corresponding list of local and shared ids.

**Algorithm 1** Neko's overlapped gather-scatter kernel.

1: $S \leftarrow \emptyset$
2: **for** $i = 1, 2, \ldots, m$ **do**
3:     Post non-blocking receive on $buf(i)$
4:     $S \leftarrow S \cup i$
5: **end for**
6: $v \leftarrow gather(shared_{id}, u_L)$
7: Post non-blocking sends of $v$
8: $v \leftarrow gather(local_{id}, u_L)$
9: $w_L \leftarrow scatter(local_{id}, v)$
10: **while** $S \neq \emptyset$ **do**
11:     **for all** $j \in S$ **do**
12:         **if** non-blocking receive $j$ has completed **then**
13:             $v \leftarrow gather(shared_{id}, buf(j))$
14:             $S \leftarrow S \setminus j$
15:         **end if**
16:     **end for**
17: **end while**
18: $w_L \leftarrow scatter(shared_{id}, v)$

Neko's overlapping gather-scatter kernel for performing $w_L = QQ^T u_L$ is given in Algorithm 1. First, all non-blocking receives (`MPI_Irecv`) for shared ids are posted. The shared ids are gathered into a buffer $v$ and transmitted to neighbours sharing the same ids using non-blocking send (`MPI_Isend`) operations. During the non-blocking communication, the local ids are gathered from $u_L$ into a buffer $v$, the same buffer as for the shared ids. The buffer now contains all contributions from both shared and local elements (since both gather operations has been performed) for global ids that are not shared with another PE, and they can be scattered into the corresponding places in the output vector $w_L$ (for the local elements). Once the local operation has completed, a loop polls each of the posted non-blocking receives until all have completed. As data is received in the loop, it is directly gathered into the buffer $v$. Finally, with all outstanding receives completed, the shared ids are scattered back into the output vector $w_L$.

Furthermore, global ids are classified as injective or non-injective, depending on the mesh topology. Non-injective global ids are points located in corners and edges (in three dimensions) and could have an arbitrary number of neighbours to consider while performing the gather-scatter operation. Injective ids are the points on the interior of an edge or face (in three dimensions) with only a single neighbour to operate on. The non-injective ids are stored in variable-length blocks, similar to `gslib`, and the injective part is stored as a contiguous block of sorted tuples, allowing for efficient use of wide SIMD units or vector registers [29].

## 4 NUMERICAL METHOD

Neko advances the incompressible Navier-Stokes equations in time,

$$\frac{\partial u}{\partial t} + (u \cdot \nabla) u = -\nabla p + \frac{1}{Re} \nabla^2 u + f,$$
$$\nabla \cdot u = 0,$$

where $u$ is the velocity, $p$ the pressure, $f$ a volume force and the Reynolds number $Re = UL/\nu$, with the reference velocity and length $U$ and $L$ and kinematic viscosity $\nu$. The solver is based on conformal function spaces for both the pressure and momentum equation based on schemes developed by Orszag, Israeli, Deville, Karniadakis and Tomboulides [30, 46, 59], referred to as the $\mathbb{P}_N - \mathbb{P}_N$ method in Nek5000. Time integration is performed using an implicit-explicit scheme based on backward differentiation and $k$-th order extrapolation,

$$\sum_{j=0}^{k} \frac{b_j}{dt} u^{n-j} = -\nabla p^n + \frac{1}{Re} \nabla^2 u^n + \sum_{j=1}^{k} a_j \left( u^{n-j} \cdot \nabla u^{n-j} + f^n \right),$$

with $b_k$ and $a_k$ coefficients of the implicit-explicit scheme. Thus the discrete system to solve for a time-step $n$ becomes,

$$\Delta p^n = \sum_{j=1}^{k} a_j \left( u^{n-j} \cdot \nabla u^{n-j} + f^n \right),$$

$$\frac{1}{Re} \Delta u^n - \frac{b_0}{dt} u^n = \nabla p^n + \sum_{j=1}^{k} \left( \frac{b_j}{dt} u^{n-j} + a_j \left( u^{n-j} \cdot \nabla u^{n-j} + f^n \right) \right).$$

| Solver | Pressure | | Projections |
|--------|----------|---|-------------|
| | Krylov Solver | Coarse grid solver | |
| Neko | GMRES | CG | 20 |
| Nek5000 | GMRES | XXT | 8 |
| NekRS | Flex-CG | HYPRE-AMG | 8 |

Table 2: Solver characteristics of the three different codes.

In each time step, we solve one Poisson equation to obtain the pressure using extrapolated velocities on the boundaries, followed by a Helmholtz equation for each velocity components. All systems are solved for using iterative Krylov subspace methods. A preconditioned Conjugate Gradient (CG) method is used for velocity with a block Jacobi preconditioner. For the Poisson equation, we use a preconditioned Generalised Minimal Residual Method (GMRES). The Poisson equation is the most challenging part to solve in the system, and we, therefore, use similar projections techniques as in Nek5000 [20] to accelerate convergence by storing a set of previous solutions. Nevertheless, an efficient, scalable preconditioner is essential and we here employ a similar two-level additive overlapping Schwarz method as in Nek5000 [19, 22], given by

$$M_0^{-1} = R_0^T A_0^{-1} R_0 + \sum_{k=1}^{K} R_k^T \tilde{A}_k^{-1} R_k,$$

for a general $k$-level formulation, where $R_k$ and $R_k^T$ are the restriction and prolongation operators to move between different grid levels. The coarse grid (on linear elements) is solved for using an approximate Krylov solver, in essence performing few ($\approx 10$) CG iterations. Possible negative effects of using a less accurate coarse grid solve are mitigated via an increased projection space.

## 5 EXPERIMENTAL SETUP

For the experiments in this paper, we use Nek5000 version 19.0, NekRS version 20.1 (both codes cloned on 9th of March 2021), and a pre-release of Neko. Since each code runs with various options, impacting both performance and numerical stability, we choose options to give each code the optimal performance to the best of our knowledge. The options that differ for each code are given in Table 2.All the codes use a two-level additive Schwarz for the pressure preconditioner but different coarse grid solvers. For all

Niclas Jansson, Martin Karp, Artur Podobas, Stefano Markidis, and Philipp Schlatter

Listing 2: Outline of the matrix-vector product benchmark.
```
program bench
  use neko
  type(space_t) :: Xh
  type(coef_t) :: coef
  type(mesh_t) :: msh
  type(ax_gpu_t) :: kernel ! extends ax_t
  ! initialise Xh, coef, msh via calls to libneko
  ...
  ! Compute kernel on a given backend
  do i = 1, n
    call kernel%compute(w, u, coef, msh, Xh)
  end do
end program bench
```

codes, we use a BDF3 scheme for the time-stepping, the Pn-Pn method without dealiasing and a CG solver with a Block-Jacobi preconditioner to compute the velocity. One improved feature in NekRS, not available in Nek5000 or Neko, is the use of extrapolation to make an initial guess for the velocity.

We run all tests on a 1676 node Cray XC40, each having 64GB of RAM and two 16 core Intel E5-2698v3 running at 2.3 GHz. The Intel compiler version 19.1.1.127 are used to compile both Nek5000 and Neko, with optimisation flag -O3 for Neko. NekRS requires the GNU compilers; the code was built using GCC version 10.1.0. For both NekRS and Nek5000 the flags used were the standard build setup.

Additionally, we run Neko on a heterogeneous compute cluster with 64 NEC A300-8 nodes for the performance evaluation. Each equipped with eight 1.4GHz NEC SX-Aurora TSUBASA Type 10B Vector Engines, with 48GB of HBM2 memory and eight cores each. The NEC Fortran compiler version 3.1.1 is used with optimisation flags -O3 -finline-functions and all experiments were using the native execution mode of the Vector Engine. Portability experiments also included an Nvidia A100 GPU using GNU compilers version 10.2.0 and CUDA version 11.1, an AMD MI100 GPU using Cray Compiling Environment 11.0.2 and AMD ROCm version 4.1.0, and an Intel Stratix 10 GX2800 FPGA using GNU compilers 10.2.0 and Intel OpenCL SDK version 20.2 (Quartus Prime v19.4).

## 6 PORTABILITY

One of the more salient strengths of Neko is portability. Today, Neko fosters the use of accelerators in HPC by supporting emerging systems such as Field-Programmable Gate Arrays (FPGAs), Graphical Processing Units (GPUs), or Vector processors. Internally, Neko facilitates the use of accelerators by overloading the abstract target type with an accelerator-specific version inside of libneko.

Consider, for example, the computation of a matrix-vector product of the unassembled stiffness matrix of the Poisson operator (9). This computation is among the more computationally demanding functions in CFD-code, making it a suitable candidate for accelerators. Leveraging Neko's infrastructure for accelerators, we ported the candidate (internally represented by the abstract type ax_t) to several different architectures, including the Stratix 10 FPGA, the Nvidia A100, the recent AMD MI100 GPU, as well as the SX-Aurora system. Such diverse support for heterogeneity (FPGAs, GPUs, CPUs, Vector cards) is hard to achieve without the necessary infrastructure provided by Neko.

To demonstrate the performance, we executed the accelerated portion on the accelerators as well (to provide contrast) on a single socket (16 cores) of a Cray XC40 node. We measure the time to compute the platform-specific (derived) ax_t. All platform-specific code, such as memory allocation on the accelerator, data transfer routines, were all isolated to the derived type and invoked implicitly via the call to compute in Listing 2.

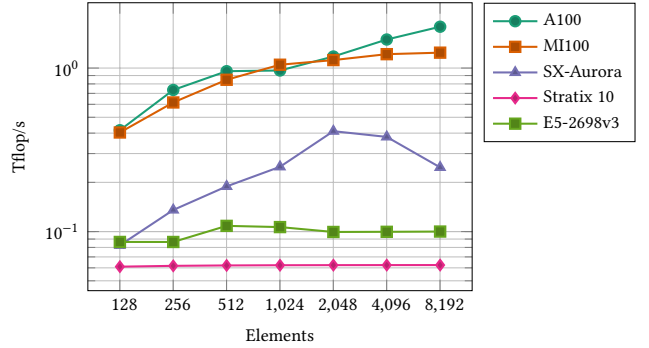

Figure 2: .Floating point operations per second (flop/s) for the matrix-vector product benchmark on various platforms.

Figure 2 compares the average floating point operations per second for the matrix-vector benchmark when executed on all the different platforms using different meshes ranging from 128 elements up to 8192 elements with ten quadrature points in each direction. In general, we can obtain good performance on most platforms, achieving more than 10% of theoretical double precision peak performance, demonstrating that Neko is portable with low overhead costs in the abstraction layer. Details of the FPGA implementation is given in [32], the SX-Aurora backend [29] and the GPU backend follows the tuned kernels by Świrydowicz et. al. [58].

## 7 VERIFICATION

In this section, we present our verification of Neko and compare it to both Nek5000 as well as NekRS. For the verification we employ the Taylor-Green vortex (TGV) test case with Reynolds number ($Re$) equal to 1,600. This test case has been studied extensively as a benchmark to assess the accuracy of higher order CFD methods [7, 62, 63]. The case setup consists of a cube with side length $2\pi$ and periodic boundaries. The initial conditions at $t = 0$ for the Taylor-Green vortex for this domain are

$$\begin{aligned} u_x(x,y,z) &= \sin(x)\cos(y)\cos(z), \\ u_y(x,y,z) &= \sin(x)\cos(y)\cos(z), \\ u_z(x,y,z) &= 0. \end{aligned} \quad (11)$$

For our particular setup we use a time step dt = $5 \cdot 10^{-4}$, leading to a maximal CFL number of 0.20.

One key aspect of the Taylor-Green vortex is that its characteristics and dissipation of energy are controlled by the computational Reynolds number given the above analytical initial conditions. This means it is possible to compare the precision of a solver by measuring e.g. the enstrophy and the kinetic energy of the simulation at any given time point. The enstrophy $\mathcal{E}$ is a property of turbulent flow that is closely related with the dissipation of kinetic energy in the fluid and is therefore very sensitive to the numerical dissipation and accuracy of the solver. For an incompressible fluid the enstrophy can be computed as the volume integral of the vorticity $\omega$ squared $\mathcal{E}(\omega) = \int_\Omega |\omega|^2 dx$.

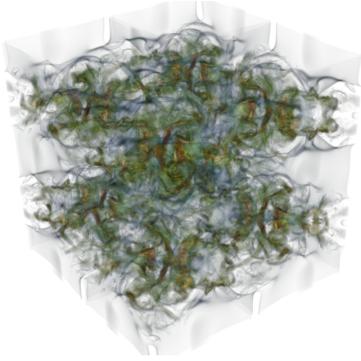

Figure 3: **Volume rendering of the velocity magnitude of the turbulent flow in the Taylor-Green vortex at** $Re = 5000$.

Therefore, by computing the enstrophy, and relating it to the dissipation of kinetic energy as $\frac{2}{Re}\mathcal{E}$ which holds for incompressible flow, we can relate the result to readily available DNS data [63] and assess the accuracy of the three solvers. We verify and validate all the solvers by running a simulation on a mesh with $64^3$ elements and a polynomial degree of 7 leading to a total number of $512^3$ nodal points and resolving $448^3$ grid points. The reference DNS data has a resolution of $512^3$ and was obtained with a dealiased pseudo-spectral method [63].

## 7.1 Verification Results

As can be seen in both Figure 4 and 5 all the solvers match the reference DNS data closely, all within 0.25% of the reference DNS solution with regards to the enstrophy. Comparing the three different SEM solvers to each other, the difference is even smaller. They all match within a tolerance of $10^{-6}$ for the enstrophy, between Nek5000 and Neko , the largest difference being of order $10^{-11}$. Considering that Nek5000 has been widely used for decades with satisfactory results, it is evident that both NekRS and Neko also live up to the high standards of their predecessor.

As for the accuracy of the SEM itself, the error for all methods is relatively small. Since our simulations' resolution is also close to reference DNS data, we also expect this to be the case. Compared to other studies where they have used high-order methods, at lower resolution though, we achieve higher accuracy [63].

## 8 PERFORMANCE MODEL

Going forwards, we will provide a model for the expected performance of SEM with a particular focus on Neko. One of the most widely used tools to evaluate the bounds of an application is the Roofline model [64]. Since its proposal, it has been extended in several ways by taking into account different levels of cache, memory systems and has been integrated into a wide range of profiling tools [5]. It can give tremendous insight into which kernels require code optimization and what are the performance bottlenecks for specific kernels in a large scale application [34]. Nek5000 has previously been shown to be memory bound, and we will therefore use the roofline model to relate the performance of each single kernel in Neko to the peak performance and memory bandwidth. Extending this to the entire solver is a natural extension of previous work focusing on the computationally most intensive kernels that perform the tensors operations in spectral methods [31, 32, 58]. In addition to this, we apply the linear communication model (also known as postal or Hockney model) for Neko, which has previously been used to model the scaling limits of Nek5000 [21]. In this section, we present the resulting model by combining these approaches for Neko. In later sections, we then use the model to evaluate our performance results and discuss the performance on future architectures.

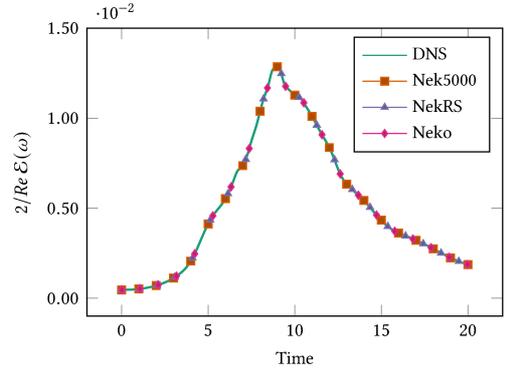

Figure 4: **The rate of dissipation of kinetic energy for the three different solvers. The results closely overlap.**

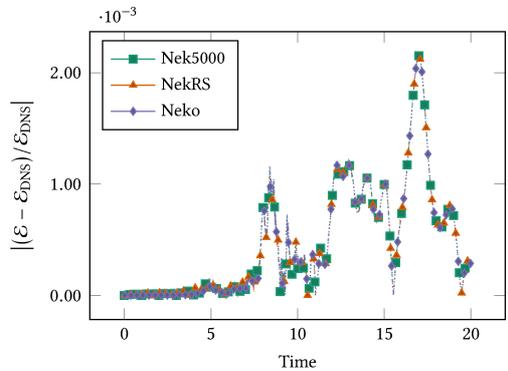

Figure 5: **The relative error of Neko, Nek5000 and NekRS compared to the reference DNS data with regards to the enstrophy as a function of time.**

In order to model the performance of Neko, we focus on the total time $T$ to compute one simulation time step. Creating a performance model is always a question of selecting the correct problem specific and hardware parameters. While a simple model is easy to generalize, it might not be accurate and while a complex model with many parameters can be incredibly detailed, the risk of overfitting is also large. In our discussion we will base the entire model on the following parameters: the polynomial order $N$ and total number of elements $E$, the total number of points $n = EN^3$ and the problem size, the number of GLL-points, $n_p = E(N + 1)^3$, the number of pressure iterations $j$ and velocity iterations $i$, the dimension of our projection space $m$, the global memory bandwidth $\beta$, the peak performance per PE $\pi$, the network latency $\alpha^*$ and bandwidth $\beta^*$, and finally the number of PEs $P$. With these parameters, the only empirical measurements we need for a given platform is the bandwidth, peak performance and the specs of the interconnect. To determine these values, we use micro-benchmarks, such as the STREAM (the triad kernel specifically) [39] and Pingpong tests. Since the parameters are quite general, we believe a similar model can be applied to other systems with accurate results.

We start the analysis by splitting the total time into two separate parts, the arithmetic and communication times $T = T_a + T_c$. While an argument can be made that $T = \max\{T_a, T_c\}$, in the case of Neko, the time waiting for MPI-collectives such as allreduce are not overlapped with computation. In general, the PE needs the computed value from the reduction in the following arithmetic operations and the time is more accurately modeled as $T = T_a + T_c$.

As for the arithmetic time, we model this with a similar approach as that in the Roofline model. We will assume cold cache between different subroutines/kernels, but that any kernel in itself has perfect locality and performs on the Roofline. We would expect this behavior to closely resemble the performance when there are many points per PE, i.e. when $n_p/P$ is large. With this model, we aim to capture the current behavior of the code, but by restructuring and merging kernels, this performance can potentially be improved.

For each subroutine/kernel we assign an arithmetic cost $C_a$ according to the following

$$C_a = \{W, Q\},$$
$$W = \#\text{Adds} + \#\text{Mults} + \#\text{Divs}, \quad (12)$$
$$Q = \#\text{Loads} + \#\text{Stores},$$

where we only consider mandatory loads and stores based on the assumption of cold cache. These costs for different kernels can then easily be added together recursively. By computing the cost for low level kernels, we work our way up until an entire time step is computed. For clarity, if a subroutine CopyAdd calls two other lower level kernels Copy and Add, the cost $C_{a,\text{CopyAdd}} = C_{a,\text{Copy}} + C_{a,\text{Add}}$.

Relating the cost for time step $C_{a,\text{step}}(n_p, N, i, j, m) = W_{\text{step}} + Q_{\text{step}}$ to the time for one time step $T_{a,\text{step}}$ then becomes a procedure of relating the cost to the peak performance and global memory bandwidth

$$T_{a,\text{step}} = \max\left\{\frac{W_{\text{step}}}{\pi P}, \frac{Q_{\text{step}}}{\beta P}\right\}. \quad (13)$$

As for the scaling behavior of Neko and SEM, the postal model has previously been used to model the communication time $T_c$ for Nek5000 [21, 44]. The postal model only depends on two parameters and is commonly written as $t_c(m) = \alpha^* + \beta^* m$ where we take into account the latency of the network $\alpha^*$ in seconds and the inverse of the bandwidth $\beta^*$ in s/64-bit word to compute the time $t_c$ to send a message with $m$ 64-bit words. Previous work has applied this analysis to the CG and multigrid parts of the solver to assess the scaling limits in the context of Nek5000 [21], but we extend it to the entire solver. To model the communication, we start with the observation that the only calls in Neko that require MPI during a time step are MPI-Allreduce and the gather-scatter operation. The time for Allreduce has previously been modeled by Fischer as $t_{c,\text{Allreduce}} = 2\alpha^* \log_2(P)$ assuming a contention-free binary fan-in/fan-out. However, modeling Allreduce with only latency and bandwidth has certain limitations, in part because of its simplicity and in part because we relate the time of each reduction level to the latency directly [26, 49]. The question is if $2\alpha^*$ is an acceptable approximation for the time spent at each reduction level. While the communication cost of allreduce is $O(\log_2(P))$, the constant factor has a large impact on the actual scaling limit of these solvers. One issue is that noise can have a large impact on the performance of MPI [27]. The latency can be highly stochastic, in particular for a Cray XC40 [44]. For the Cray XC40 used in our tests, the average $\alpha^*$ was 2.6$\mu s$ while the 99th percentile takes more than 20$\mu s$. We, therefore, argue that a better approximation of the time for Allreduce, at least, in this case, is determined by the following

$$t_{c,\text{Allreduce}} = \max_{p \in \{1,\ldots,P\}} \sum_{i=1}^{\log_2(P)} \alpha_i^* \quad (14)$$

which is analogous to modeling the maximum time it takes for one PE to receive the message from the PE $\log_2(P)$ levels away. The latencies $\alpha_i^*$ are then sampled from a sample of over 10,000 measured Pingpong latencies across different node configurations. As for the gather-scatter operation, this is also modeled as,

$$t_{c,\text{gather-scatter}} = 2\beta^* n_p^{2/3} + \max_{i \in \{1,\ldots,6\}} (\alpha_{i,1}^* + \alpha_{i,2}^*) \quad (15)$$

where we assume any PEs elements are perfectly arranged in a cube and the time spent in communication is limited by the latency by the slowest the PEs neighbors. The gather-scatter corresponds to two MPI-calls, one scattering half of the external boundary to (3) neighbours and one gathering the results from neighboring the three other sides PEs. However, it should be noted that the cost of allreduce is several times higher than the gather-scatter operation on CPUs. In particular, since Neko uses CG for the coarse grid solve, the communication cost associated with Allreduce makes up the vast majority of modeled communication time.

Now, it is possible, by counting the number of calls to Allreduce and gather-scatter per time step, to obtain the communication time per step $T_{c,\text{step}}$. In total, we then obtain $T_{\text{step}} = T_{a,\text{step}} + T_{c,\text{step}}$. With this time we can relate the Neko performance of Neko to the system network and the global memory bandwidth. As CPUs have large amounts of cache though, the model will likely be beat when cache has a major impact. As for future large and relevant problems and architectures such as GPUs where the cache size is comparatively small or when $n_p/P$ is large, we anticipate that a model such as this can be applied with great effectiveness.

## 9 PERFORMANCE EVALUATION

To assess the performance of Neko, we carry out a strong scaling study. We use three test cases with varying characteristics and sizes, comparing average time per time-step when executing on a different number of processing elements. As a comparison, we execute the same experiments with Nek5000 and NekRS.

The first test case (Hemi) studies the interaction of a flat-plate boundary layer with an isolated hemispherical roughness element at $Re_\delta = 700$ similar to the classic Nek5000 case from SC99 [61] and a scaled-downed version of one of the benchmarks used in their Gordon Bell Price submission the same year [60]. We use a mesh consisting of 2,042 hexahedral elements, with ten quadrature points in each direction and compute the solution in the time interval $t = [0, 20]$ with a time-step of size $dt = 10^{-3}$, measuring the execution time per simulation time step in the interval $(0.1, 20]$. As a second test case, we consider turbulent flow in a straight pipe

Neko: A Modern, Portable, and Scalable Framework for High-Fidelity Computational Fluid Dynamics

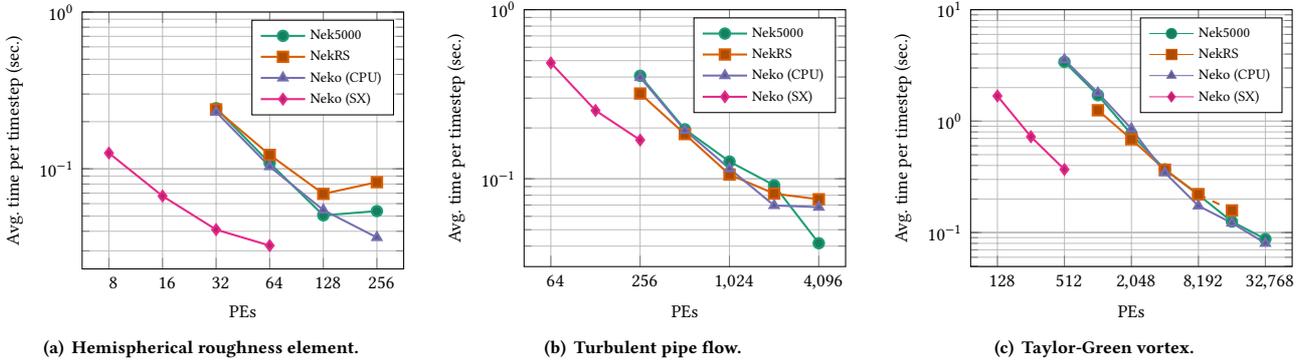

Figure 6: Average execution time per time-step for each test case when executed on the Cray XC40 (Nek5000, NekRS and Neko (CPU)) and NEC SX-Aurora TSUBASA (Neko (SX)). Dashed lines refer to data with incomplete statistics.

at a friction Reynolds number $Re_\tau = 180$. The mesh consists of 36,480 hexahedral elements, with eight quadrature points in each direction. This setup is similar to the smallest test case in Offermans et.al. [44]. In our experiments, we compute the solution in the interval $t = [0, 30]$ using a time-step of $dt = 10^{-3}$ and use the interval $(20, 30]$ for statistics. For the last test case, we use the Taylor-Green vortex (TGV) problem from Section 7, but compute at $Re = 5000$, over an interval $t = [0, 10]$ with $dt = 5 \cdot 10^{-4}$, using a hexahedral mesh with 262,144 elements and ten quadrature points in each direction. The average time per time-step is measured over the interval $(6, 10]$.

As mentioned in Section 5, we run the performance evaluation for all codes on the Cray XC40, with additional runs using Neko on the Vector Engine. For all results in 6, PE refers to a processing element on each platform: scalar cores on CPUs and vector cores on Vector Engines. Figure 6(a) compares the average execution time per simulation time-step when solving the hemispherical roughness case on one to four Cray XC40 nodes and one to eight VEs. Both Nek5000 and Neko perform equally on CPUs, and NekRS slightly worse. However, due to the small problem size, scalability is quickly lost in all codes. Albeit the lower execution time, the vector processors are also struggling due to the small problem size. For eight PEs, it achieves $\approx 4.3\%$ of the theoretical peak performances (measured via NEC's `proginf`) but at 64 PEs it has reduced down to $\approx 2.1\%$. The performance on a VE is directly related to the fraction of vector operations and their average length. The SX-Aurora has a vector length of 256, 64bit elements and utilising the full length is a must to ensure good performance. For a high enough polynomial order, most compute-intensive kernels in Neko spend $> 98\%$ of the time performing vector operations, all at full length. However, the smaller coarse grid problem reduces the average length below 100, thus the reduced performance.

Figure 6(b) compares the performance for the turbulent pipe case. With the larger problem size, we see that all three codes retain scalability for a larger number of PEs in the CPU experiments. Compared to the smaller case (Figure 6(a)), Nek5000 continues to scale well even for few elements per core in contrast to previously published results [44]. Furthermore, with the lowest polynomial order of all test cases, we obtain even smaller average vector lengths, and the VE experiments struggle throughout, reaching a sustained performance of only $\approx 1.1\%$ of peak, despite more elements per PE.

In Figure 6(c), the average execution per simulation time-step is compared for the larger scale-out Taylor-Green vortex case. We observe how the larger problem size and the 9th order polynomials scales well on the VE, with strong scalability up to the entire machine with 64 VEs (512 PEs). The average vector length is also increased compared to the other two cases, albeit less than half of the optimal length resulting in a sustained performance of $\approx 4\%$ when executing on the entire machine. This could be compared to 5% for the Nov. 2020 HPCG results[1] for the SX-Aurora. For the CPU results, the large problem size causes issues for NekRS, which can not fit the entire problem on less than 1024 PEs. Unfortunately, the problem continues for NekRS, with the solver diverging when running at 16,384 PEs. However, it managed to complete more than 95% of the run; hence we included the incomplete statistics marked as a dashed line in 6(c). Overall, given the large problem size, all codes scales well, even down to $\approx 8$ elements per PE for Nek5000 and Neko at 32,768 PEs.

## 9.1 Modeled Performance Comparison

Let us now focus on the projected performance of our model and contrast it with our actual measurements for Neko. The performance estimates from the model is based on the polynomial degree, size of projection space and number of PEs, as well as the measured average number of iterations for all cases. The results for the different cases is shown in Figure 7. Investigating the performance of all cases, we note that the modeled performance, as expected, is closest when the problem-size per PE $n_p/P$ is large, and works better overall for larger cases. This is also the domain we are most interested in when making predictions going forward. For the TGV case at $P < 2048, n_p/P > 128000$ the relative difference is less than 5%, reaching an effective bandwidth of 84GB/s compared to triad of 90GB/s. This is in accordance with our initial observations when designing the model, assuming cold cache between kernel invocations. It also supports that the kernels in Neko is performing close to the DRAM roofline. Additionally, it is evident that the effect of caches grows considerably when decreasing $n/P$, leading to a maximal difference at $n_p/P = 64000$ for the TGV case and $n_p/P = 72960, P = 256$ for the pipe. Considering the L3-cache size of $2 \cdot 40$ MB per node, this means that the it can fit more than 4.5 arrays of length 70000 across all cores in L3, implying that the possible reuse of arrays is extensive. However, even in the domain where the effects of caches are large, the modeled performance is still within 20% of the experimental results for the Pipe and TGV. As for the Hemi case, the effect of caches becomes even more prominent as already at $P = 32$, $n_p/P = 63000$, meaning that a large part of the problem can fit in LLC. As $P$ increases, the cache is large enough to even fit all arrays used in one timestep. We could for this case make the argument that the model parameter $\beta$ should be replaced with the bandwidth to L3 cache instead. Hence, we expect that Hemi should differ substantially from the model, which it does. The model overestimates the time by a factor 2. It is evident that extending the model, or rather restructuring the code, to better utilize caches and obtain a higher operational intensity for larger $n_p/P$ is a future development that we anticipate would yield both higher performance and more accurate predictions for smaller cases.

---
[1] https://www.hpcg-benchmark.org/

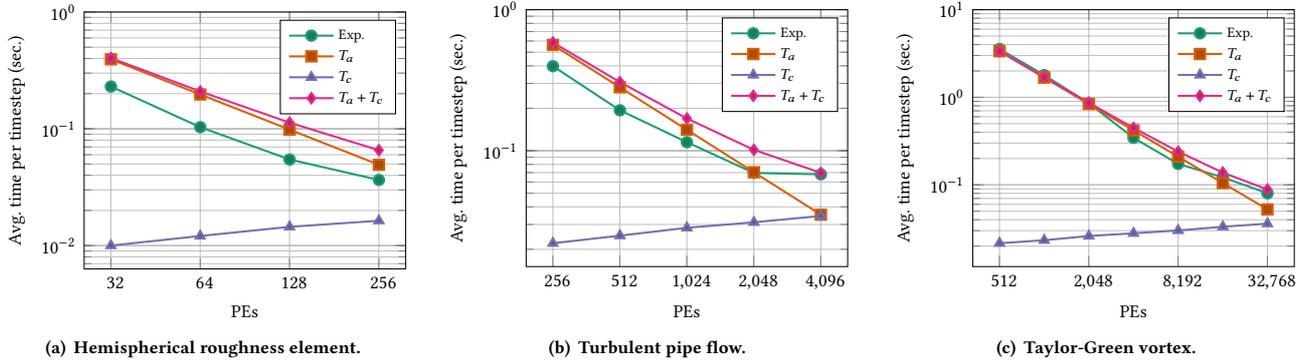

(a) **Hemispherical roughness element.**    (b) **Turbulent pipe flow.**    (c) **Taylor-Green vortex.**

Figure 7: Modeled performance compared with experimental measurements for Neko.

As for the modeled communication of Neko, this can be more thoroughly analyzed and improved. While the asymptotic communication complexity of Allreduce is $O(\log_2(P))$, applying this to determine any exact time spent in communication requires another level of complexity. In our approach, we try to remedy this to some extent by taking the randomness of the latencies into account. Overall, this approach has been successful. The scaling limit for the two larger cases correspond well to when $T_c$ becomes the dominant factor. What is clear, however, is that the model implies that the cost of Allreduce compared to gather-scatter is large. Because of the many reductions in CG this implies that moving to another coarse grid solver can improve the strong scalability of the solver.

Moving to the vector processor, we also see how valuable our model is for evaluating the solver on a new platform. Considering the large global memory bandwidth of 1TB of the vector processor, the performance model indicates that the performance would be approximately five times higher than measured for TGV and up to 10x for the two smaller cases. Considering the setup of the current pressure preconditioner, we are surprised that the performance still is so competitive compared to CPUs. With the insights from the performance model we therefore see that more tuning, algorithm development, and bottleneck analysis is required for the Vector Processor. We anticipate that similar algorithm developments, in particular for the preconditioner, will be necessary for GPUs.

## 10 PROJECTION AND FUTURE OUTLOOK

We end the study by investigating future and hypothetical HPC machine performance on Neko. The International Roadmap for Devices and Systems (IRDS) [25] was established in 2016 and aims at identifying trends related to (amongst others) computing. We extracted IRDS predictions on external memory bandwidth (HBM), core count, interconnect bandwidth/latency and used them to drive our performance model. Our prediction baseline was the Fujitsu A64FX processor, as it is currently the highest-performant general-purpose processor in existence[54]. Next, we synthesize several different configurations: `A64FX-Aries` uses the same network as the XC40 supercomputer but where we replaced Haswell processors with A64FXs', `A64FX-Tofu` is our configuration that mimics the Fugaku supercomputer, `Future (2023)` is a hypothetical Fugaku computer built on IRDS predictions for next year (2023), `Future(2027)` is a hypothetical HPC system built five years from now, and `Far Future` is an HPC machine based on what IRDS predicts that we will reach in the far future. This projection assumes the TGV case at $Re = 5000$ presented earlier.

The results are shown in Figure 8, which presents the predicted performance (per timestep) as a function of the number of cores. Note that the number of nodes used in this projection is the same, but assumes that nodes become more powerful due to improvements in fabrication and technology scaling (Moore's law, assuming it holds that far); we also included the empirical performance measured on our HPC system in order to provide a contrast to the expected performance. Overall, we can expect significant speedup when using Neko on Fugaku already, which should yield a factor 12x faster. Going further, the external memory bandwidth increases, and so does our expected performance, before plateauing already in 2023 (according to IRDS), where a peak performance of 18.6x over

Neko: A Modern, Portable, and Scalable Framework for High-Fidelity Computational Fluid Dynamics

today is obtained. Going further into the future, assuming we can push network latencies further, could – in theory – yield a speedup over 28.77x over today.

These results and preliminary projects suggest that large scale CFD simulations have ample opportunities to leverage the increased compute performance and memory bandwidth that future architecture will facilitate. This, of course, assumes a framework such as Neko that can cater to changes in the hardware. Obtaining a 28.77x performance improvement when not at the strong scale limit will lead to larger and faster simulations and could even facilitate the ensemble of simulations, directly leading to breakthroughs in the field (see CFD challenges for 2030 [56]). A final remark on architectures: in this projection, we have used IRDS data that seems to discard some of the more exciting emerging technology, such as 3D stacking [4] of silicon dies or Post-Moore technology such as CGRAs [50], which potentially could leave an even larger mark of performance in future systems.

What remains is the issue of strong scaling. While future memory systems such as HBM can yield tremendous performance, it will also more quickly hit the limit of the network. However, in the Far Future case, we could still potentially have a 20x better performance compared to our Haswell experiments at the scaling limit, assuming stable latency of 250ns. However, any noise could impact this tremendously. What is evident from our performance analysis though, is that the scaling limit is reached asymptotically at the point where $T_a = T_c$ or as we model it, #Allreduce $\alpha^* \log_2(P) \approx \frac{Q_{\text{step}}}{\beta P}$. It is therefore clear that for incredibly performant architectures, if we cannot reduce the network latency ( and noise) further, we will not be able to continue to achieve strong scaling. An interesting note is that as each node becomes more performant, the factor $\log_2(P)$ potentially decreases, implying that fewer, but higher performing nodes might help in this regard. Even if the strong scale limit of the number of points per PE, $n/P$, before we stop scaling is higher for certain architectures, as long as they can make the same computation in the same time, but on fewer nodes, this would in theory yield higher performance since $\log_2(P)$ is smaller. In the end, we are interested in minimizing the time to solution $T$, and therefore, we argue that rather than aiming for a low $n/P$, the more important measure for strong scalability is therefore to have a low $n/(P\beta)$ which would, according to the model be a better indicator of the runtime. The importance of strong scalability for these types of computations is also discussed by Fischer et al. in [18] where they stress the importance of strong scaling for several similar state of the art codes. We add to their analysis by connecting hardware parameters $\beta$ and $\alpha^*$ explicitly to the time necessary to compute one time step, relating this to the scaling and applying this to an entire solver, not only bake-off problems, with good results.

Another way forward is to also change the algorithm rather than improving the hardware, which in turn would change the arithmetic cost of our solver. Since the bandwidth is the limiting factor, efforts to restructure the code to more efficiently use caches and limit the number of accesses to global memory will therefore be essential to obtain high performance on GPUs, Vector Processors and also AMD CPUs that have a relatively low DRAM bandwidth per core. As for the scaling limit, the primary way to improve this would be to reduce the number of reductions, which could be addressed with non-blocking or pipelined Krylov solvers. As for other algorithmic improvements, we have in this paper shown that a simple CG for the coarse grid solve can give satisfactory results. Considering that the coarse grid is one of the most expensive parts of the solver, maybe we should reconsider old truths regarding the preconditioner? Moving to other precision formats or using other novel approaches to reduce the communication might be required to yield satisfactory performance and scaling capabilities on the next generation of supercomputers.

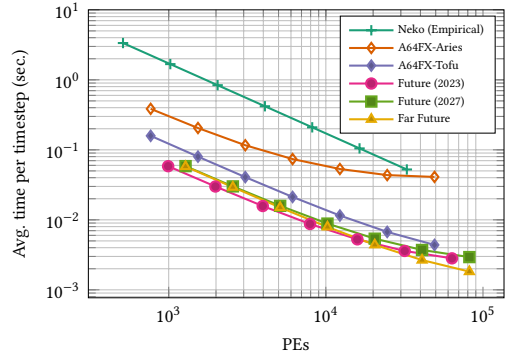

Figure 8: **Performance projections using future (hypothetical) HPC machines driven by IRDS data, showing how Neko could nearly speedup 30x in the future over today.**

## 11 RELATED WORK

Performance portability for High-order/spectral finite elements on contemporary architectures is an active topic pursued by several groups. The Center for Efficient Exascale Discretizations [36] purse this with OCCA for (among other codes) Nek5000, NekRS and MFEM [42]. The FEM library deal.II [1] provides a SIMD abstraction for their matrix-free formulation [35], while the SEM framework Nektar++ [6] is exploring various options such as Kokkos and OpenMP [14]. A performance study including several of the mentioned codes was recently published by Fischer et. al. [18].

## 12 CONCLUSION

In this paper, we have introduced Neko, a modern, high performance and flexible framework for spectral element based fluid dynamics. We have obtained similar levels of performance compared to two other state of the art solvers, Nek5000 and NekRS and have shown that having a simple CG solver for the coarse problem can measure up to more complex preconditioners such as XXT or AMG. In addition to this, we have put forth a performance model for the *entire* solver that matches the experimental results. We then used this model to reason around the performance impact of future architectural and algorithmic improvements for this family of solvers.

Overall, we see that the future development of the Spectral Element Method has many paths forward, and we intend to carefully evaluate several of these with Neko, which offers us the performance, scalability and precision required for large production runs, but with the new opportunity to easily develop and try new architectures, methods and algorithms.

## ACKNOWLEDGMENTS

This work was supported by the European Commission Horizon 2020 project grants "EXCELLERAT: The European Centre of Excellence for Engineering Applications" (grant reference 823691) and "EPiGRAM-HS: Exascale Programming Models for Heterogeneous Systems" (grant reference 801039) and the Swedish Research Council project grant "Efficient Algorithms for Exascale Computational Fluid Dynamics" (grant reference 2019-04723). Financial support from the SeRC Exascale Simulation Software Initiative (SESSI) is also gratefully acknowledged. The experiments were performed on resources provided by Höchstleistungsrechenzentrum Stuttgart (HLRS) and the Swedish National Infrastructure for Computing (SNIC) at PDC Center for High Performance Computing.